\documentclass[doublespacing]{elsart}

\usepackage{epsfig}
\usepackage{amssymb}

\begin{document}
\begin{frontmatter}

\title{\bf Long term memories of developed and emerging markets: using the scaling analysis to characterize their stage of development}
\author[label1,label2]{T. Di Matteo},
\author[label2]{T. Aste},
\author[label3,cor1]{M. M. Dacorogna}
\address[label1]{INFM - Dipartimento di Fisica "E. R. Caianiello",
Universit\`a degli Studi di Salerno, 84081 Baronissi (SA),
Italy.}
\address[label2]{Applied Mathematics, Research School of Physical Sciences, Australian National University, 0200 Canberra, Australia.}
\corauth[cor1]{Corresponding author:Tel: +41 1 6399760, fax: +41 1 6399961.}
\ead{michel.dacorogna@converium.com}
\address[label3]{Converium Ltd, General Guisan - Quai 26, 8022 Zurich, Switzerland.}

\begin{abstract}
The scaling properties encompass in a simple analysis many of the volatility characteristics of financial markets. That is why we use them to probe the different degree of markets development. We empirically study the scaling properties of daily Foreign Exchange rates, Stock Market indices and fixed income instruments by using the generalized Hurst approach. We show that the scaling exponents are associated with characteristics of the specific markets and can be used to differentiate markets in their stage of development. The robustness of the results is tested by both Monte-Carlo studies and a computation of the scaling in the frequency-domain.
\end{abstract}

\end{frontmatter}

\newpage

\section{Introduction}
\label{s.Introduction}
In a recent book \cite{HFIntro}, the hypothesis of heterogeneous market agents was developed and backed by empirical evidences. According to this view, the agents are essentially distinguished by the frequency at which they operate in the market. The scaling analysis, which looks at the volatility of returns measured at different time intervals, is a parsimonious way of assessing the relative impact of these heterogeneous agents on price movements. Viewing the market efficiency as the result of the interaction of ¨these agents \cite{MMD.2001-01-01}, brings naturally to think that it is the presence of many different agents that would characterize a mature market, while the absence of some type of agents should be a feature of less developed markets. Such a fact should then reflect in the measured scaling exponents. The study of the scaling behaviors must therefore be an ideal candidate to characterize markets. 

For institutional investors, a correct assessment of markets is very important to determine the optimal investment strategy. It is common practice to replicate an index when investing in well developed and liquid markets. Such a strategy minimizes the costs and allow the investor to fully profit from the positive developments of the economy while controlling the risk through the long experience and the high liquidity of these markets. When it comes to emerging markets, it is also clear that the stock indices do not fully represent the underlying economies. Despite its higher costs, an active management strategy is required to control the risks and fully benefit from the oportunities offered by these markets. The differentiation between markets is clear for the extreme cases: New York stock exchange and the Brazilian or Russian stock exchange. The problem lies for all those in between: Hungary, Mexico, Singapore and others. For those markets a way to clarify the issue will help decide on the best way to invest assets. 

The purpose of this article is to report on the identification of a strong relation between the scaling exponent and the development stage of the market. This conclusion is backed by a wide and unique empirical analysis of several financial markets (32 Stock market indices, 29 Foreign exchange rates and 28 fixed income instruments) at different development stage: mature and liquid markets, emerging and less liquid markets. Furthermore, the robustness and the reliability of the method is extensively tested through several numerical tests, Monte Carlo simulations with a variety of random generators and the comparison with results obtained from a frequency domain computation of related exponents.

The scaling concept has its origin in physics but is increasingly applied outside its traditional domain \cite{UAM.1990-01-01} \cite{HFIntro}. In the recent years, its application to financial markets, initiated by Mandelbrot in the 1960~\cite{BBM.1963-01-01,FracScaFin}, has largely increased also in consequence of the abundance of available data \cite{UAM.1990-01-01}. Two types of scaling behaviors are studied in the finance literature:
\begin{enumerate}
\item The behavior of some forms of volatility measure (variance of returns, absolute value of returns) as a function of the \emph{time interval} on which the returns are measured. (This study will lead to the estimation of a \emph{scaling exponent} related to the Hurst exponent.)
\item The behavior of the tails of the distribution of returns as a function of \emph{the size of the movement} but keeping the time interval of the returns constant. (This will lead to the estimation of the \emph{tail index} of the distribution \cite{HFIntro}.)
\end{enumerate}
Although related, these two analysis lead to different quantities and should not be confused as it is often the case in the literature as can be seen in the papers and debate published in the November 2001 issue of Quantitative Finance \cite{BLB.2001-01-01} \cite{TLU.2001-01-01} \cite{BBM.2001-01-01}. For more explanations about this and the relation between the two quantities, the reader is referred to the excellent paper by \cite{PAG.1998-01-01}. In this study, we are interested in the first type of analysis. Until now, most of the work has concentrated in studies of particular markets: Foreign Exchange \cite{UAM.1990-01-01} \cite{HFIntro} \cite{FCO.2001-01-01}, US Stock Market (Dow Jones) \cite{RNM.1995-01-01} or Fixed Income \cite{GBA.1999-01-01}. These studies showed that empirical scaling laws hold in all these markets and for a large range of frequencies: from few minutes to few months.

Recently, a controversy has erupted between \cite{BLB.2001-01-01} on one side and \cite{BBM.2001-01-01,EHS.2001-01-01} on the other side with somewhere in the middle \cite{TLU.2001-01-01} to know if the processes that describe financial data are truly scaling or simply an artifact of the data. Moreover, these papers propose new scaling models or empirical analysis that better describe empirical evidences and one could add to these \cite{JPB.1999-01-01}. It should be however noted that - as underlined by Stanley et al. \cite{Stanle96} - in statistical physics, when a large number of microscopic elements interact without characteristic scale, universal macroscopic scaling laws may be obtained independently of the microscopic details.

Here we address the question of the scaling properties of financial time series from another angle. We are not interested in fitting a new model but want to gather empirical evidences by analyzing daily data (described in Section~\ref{s.data}). With the same methodology, we study very developed as well as emerging markets in order to see if the scaling properties differ between the two and if they can serve to characterize and measure the development of the market. Here the scaling law is not used to conclude anything on the theoretical process but to the contrary we use it as a ``stylized fact'' that any theoretical model should also reproduce. Our purpose is to show how a relatively simple statistics gives us indications on the market characteristics, very much along the lines of the review paper by Brock \cite{WAB.2000-01-02}. In Section~\ref{s.Scalinganalysis}, we recall the theoretical framework and in Section~\ref{s.Hurstgeneralized} we introduce the generalized Hurst exponents analysis. The metodology is described in Section~\ref{s.metodology}. In Section~\ref{s.res} the generalized Hurst exponents results and their temporal stability check are presented. In Section~\ref{s.frescaling} we compute the scaling exponents in the frequency domain and we compare the scaling spectral exponents and the Hurst exponents. In Section~\ref{s.monte} a Monte Carlo simulation is presented. Finally some conclusions are given in Section~\ref{s.conclusion}.

\section{Data Description and Studied Markets}
\label{s.data}
We study several financial markets which are at different development stage: mature and liquid markets, emerging and less liquid markets. Moreover, we choose markets that deal with different instruments: equities, foreign exchange rates, fixed income futures. In particular, the data that we analyze are: twenty nine Foreign Exchange rates (FX) (see Table~\ref{t.1}), thirty two Stock Market indices (SM) (see Table~\ref{t.2}), Treasury rates corresponding to twelve different maturity dates (TR) (see Table~\ref{t.3}) and Eurodollar rates having maturity dates ranging from $3$ months to $4$ years (ER) (see Table~\ref{t.4}). Hereafter we give a brief description of the time-series studied in this paper.

\begin{description}
\item [ FX:] The Foreign Exchange rates (Table~\ref{t.1}) are 29 daily spot rates of major currencies against the U.S. dollar. The time series that we study go from $1990$ to $2001$ and $1993$ to $2001$. These rates have been certified by the Federal Reserve Bank of New York for customs purposes. The data are noon buying rates in New York for cable transfers payable in the listed currencies. These rates are also those required by the Securities and Exchange Commission (SEC) for the integrated disclosure system for foreign private issuers. The information is based on data collected by the Federal Reserve Bank of New York from a sample of market participants.
\item [ SM:] The Stock Market indices (reported in Table~\ref{t.2}) are 32 of the major indices of both very developed markets like the US or European markets and emerging markets. These daily time series range from $1990$ or $1993$ to $2001$.
\item [ TR:] The Treasury rates (Table~\ref{t.3}) are daily time series going from $1990$ to $2001$. The yields on Treasury securities at `constant maturity' are interpolated by the U.S. Treasury from the daily yield curve. This curve, which relates the yield on a security to its time to maturity, is based on the closing market bid yields on actively traded Treasury securities in the over-the-counter market. These market yields are calculated from composites of quotations obtained by the FD Bank of New York. The constant maturity yield values are read from the yield curve at fixed maturities, currently 3 and 6 months and 1, 2, 3, 5, 7, 10, and 30 years. The Treasury bill rates are based on quotes at the official close of the U. S. Government securities market for each business day. They have maturities of 3 and 6 months and 1 year.
\item [ ER:] The Eurodollar interbank interest rates (Table~\ref{t.4}) are bid rates with different maturity dates and they are daily data in the time period $1990$-$1996$~\cite{TDM.2002-01-01}.
\end{description}

\section{Theoretical Framework and background}
\label{s.Scalinganalysis}
The scaling properties in time series have been studied in the literature by means of several techniques. For the interested reader we mention here some of them such as the seminal work~\cite{HEH.1951-01-01} on rescaled range statistical analysis $R/S$ with its complement \cite{StorRes} and the modified $R/S$ analysis of~\cite{AWL.1991-01-01}, the multiaffine analysis~ \cite{CKP.1994-01-01}, the detrended fluctuation analysis (DFA)~\cite{MAU.2000-01-01}, the periodogram regression (GPH method)~\cite{JGE.1983-01-01}, the $(m,k)$-Zipf method~\cite{HumBeh}, the moving-average analysis technique \cite{ArtInv}, the Average Wavelet Coefficient Method in~\cite{WavMeth} and in~\cite{RGE.2001-01-01}, the ARFIMA estimation by exact maximum likelihood (ML) ~\cite{FBS.1992-01-01} and connection to multi-fractal/multi-affine analysis (the $q$ order height-height correlation) have been made in various papers like \cite{KIV.1999-01-01}. In the financial and economic literature, many are the proposed and used estimators for the investigation of the scaling properties. To our knowledge it doesn't exit one whose performance has no defeciencies. The use of each of the above mentioned estimators can be subject to both advantages and disadvantages. For instance, simple traditional estimators can be seriously biased. On the other hand, asymptotically unbiased estimators derived from Gaussian ML estimation are available, but these are parametric methods which require a parameterized family of model processes to be chosen a priori, and which cannot be implemented exactly in practice for large data sets due to high computational complexity and memory requirements~\cite{Phillips1999a,Phillips1999b,Phillips2001}. Analytic approximations have been suggested (Whittle estimator) but in most of the cases (see \cite{Beran1994}), computational difficulties remain, motivating a further approximation: the discretization of the frequency-domain integration. Even with all these approximations the Whittle estimator remains with a significantly high overall computational cost and problems of convergence to local minima rather than to the absolute minimum may be also encountered. 

The rescaled range statistical analysis ($R/S$ analysis) was first introduced by Hurst to describe the long-term dependence of water levels in rivers and reservoirs. It provides a sensitive method for revealing long-run correlations in random processes. This analysis can distinguish random time series from correlated time series and gives a measure of a signal ``roughness''. What mainly makes the Hurst analysis appealing is that all these information about a complex signal are contained in one parameter only: the \emph{Hurst exponent}. However, the original Hurst $R/S$ approach has problems in the presence of short memory, heteroskedasticity, multiple scale behaviors. This has been largely discussed in the literature (see for instance \cite{AWL.1991-01-01,VTE.1999-01-01}) and several alternative approaches have been proposed. The fact that the range relies on maxima and minima makes also the method error-prone to any outlier. Lo \cite{AWL.1991-01-01} suggested a modified version of the R/S analysis that can detect long-term memory in the presence of short-term dependence \cite{MJLW.1995}. The modified R/S statistic differs from the classical R/S statistic only in its denominator, adding some weights and covariance estimators to the standard deviation suggested by Newey and West \cite{NeweyWest1987}, and a truncation lag, $q$. In the modified R/S, a problem is choosing the truncation lag $q$. Andrews \cite{Andrews1991} showed that when q becomes large relative to the sample size N, the finite-sample distribution of the estimator can be radically different from its asymptotic limit. However, the value chosen for $q$ must not be too small, since the autocorrelation beyond lag $q$ may be substantial and should be included in the weighted sum. The truncation lag thus must be chosen with some consideration of the data at hand.   

In this paper we use a different and alternative method: the generalized Hurst exponent method. We choose this type of analysis precisely because it combines the sensitivity to any type of dependence in the data to a computationally straight forward and simple algorithm. The main aim of this paper is to give an estimation tool from an empirical analysis which provides a natural, unbiased, statistically and computationally efficient, estimator of the generalized Hurst exponents. 
This method, described in the following section, is first of all a tool which studies the scaling properties of the data directly via the computation of the $q$-order moments of the distribution of the increments. The $q$-order moments are much less sensitive to the outliers than the maxima/minima and different exponents $q$ are associated with different characterizations of the multiscaling complexity of the signal. In the following we show that this method is robust and it captures very well the scaling features of financial fluctuations. We show that through the use of a relatively simple statistics we give a wide view of the scaling behaviour across different markets.

\section{Generalized Hurst exponent}
\label{s.Hurstgeneralized}
The Hurst analysis examines if some statistical properties of time series $X(t)$ (with $t$=$\nu$, $2\nu$,~...,~$k\nu$,~...,~$T$) scale with the observation-period ($T$) and the time-resolution ($\nu$). Such a scaling is characterized by an exponent $H$ which is commonly associated with the long-term statistical dependence of the signal. A generalization of the approach proposed by Hurst should therefore be associated with the scaling behavior of statistically significant variables constructed from the time series. To this purpose we analyze the $q$-order moments of the distribution of the increments \cite{FracScaFin,ALB.1991-01-01} which is a good characterization of the statistical evolution of a stochastic variable $X(t)$:
\begin{equation}
\label{multi}
K_q(\tau) =
\frac{ \left< |X(t+\tau) - X(t)|^q \right> }{\left< |X(t)|^q \right> }\;\;\;,
\end{equation}
where the time-interval $\tau$ can vary between $\nu$ and $\tau_{max}$. (Note that, for $q=2$, the $K_q(\tau)$ is proportional to the autocorrelation function: $a(\tau)=\left< X(t+\tau)X(t)\right>$.)

\begin{figure}
\begin{center}
\mbox{\epsfig{file=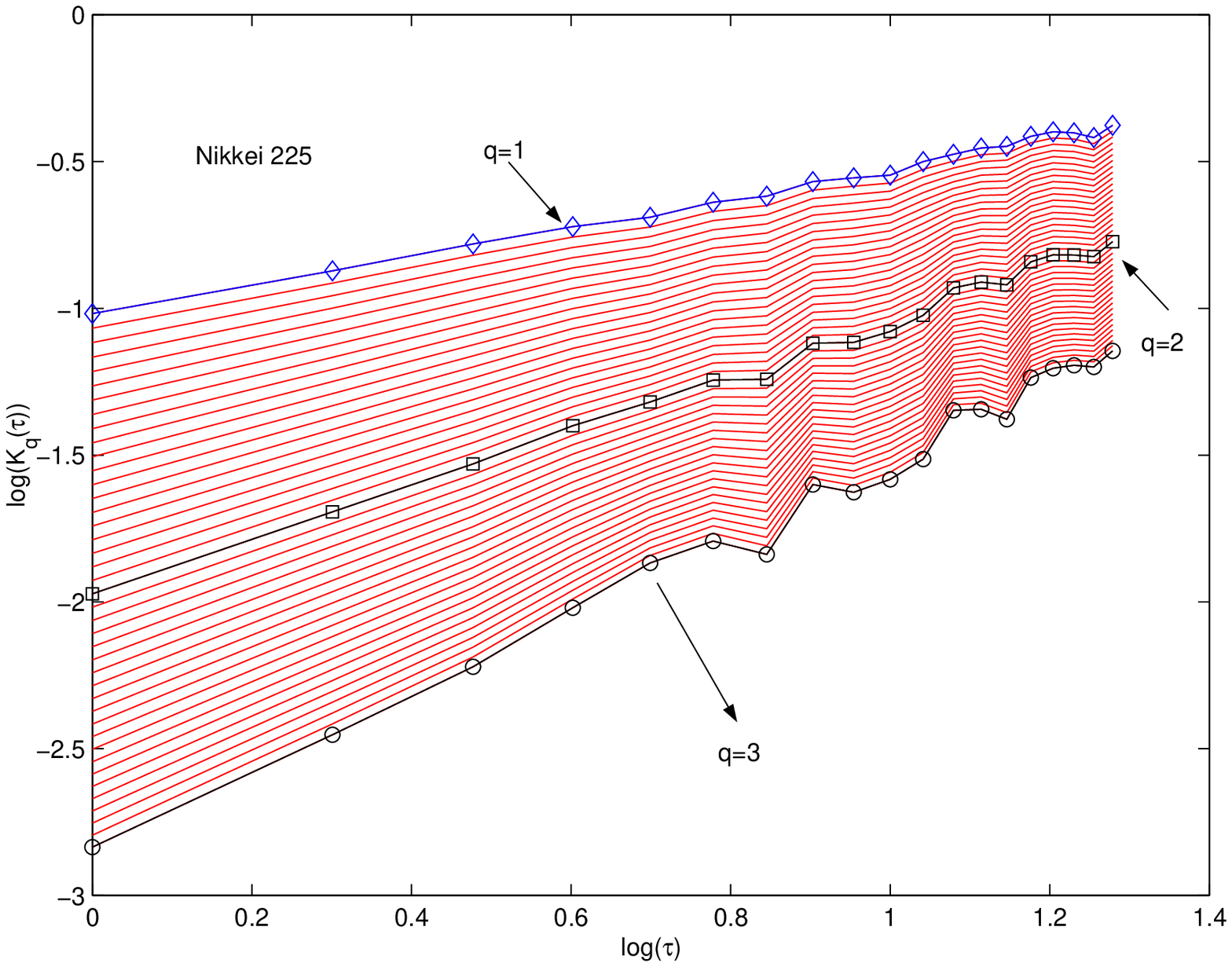,width=6.in,angle=0}}
\end{center}
\caption{$K_q(\tau)$ as a function of $\tau$ in a log-log scale for the Nikkei $225$ time series in the time period from $1990$ to $2001$ ($\tau$ varies from $1$ day to $19$ days.). Each curve corresponds to different fixed values of $q$ ranging from $q=1$ to $q=3$. In particular, the curve corresponding to $q=1$ (diamond markers), the one to $q=2$ (square markers) and the curve to $q=3$  (circle markers) are shown.}
\label{f.nik225_kq_dt}
\end{figure}

The generalized Hurst exponent $H(q)$\footnote{We use $H$ without parenthesis as the original Hurst exponent and $H(q)$ as the generalized Hurst exponent.} can be defined from the scaling behavior of $K_q(\tau)$ \cite{ALB.1991-01-01}, which can be assumed to follow the relation
\begin{equation}
\label{Hurstgen}
K_q(\tau) \sim
\left( \frac{\tau }{ \nu } \right)^{qH(q)}\;\;\;.
\end{equation}
This assumption flows naturally from the result of \cite{PAG.1998-01-01}
and has been carefully checked to be hold for the financial time series studied in this paper. For instance, in Fig.~\ref{f.nik225_kq_dt}, the scaling behaviour of $K_q(\tau)$ in agreement with Eq.~\ref{Hurstgen} is shown in the time period from $1990$ to $2001$ for Nikkei $225$. Each curve corresponds to different fixed values of $q$ ranging from $q=1$ to $q=3$, whereas $\tau$ varies from $1$ day to $19$ days. 

Within this framework, we can distinguish between two kind of processes: (i) a process where $H(q) = H$, constant independent of $q$; (ii) a process with $H(q)$ not constant. The first case is characteristic of uni-scaling or uni-fractal processes and its scaling behavior is determined from a unique constant $H$ that coincides with the Hurst exponent. This is for instance the case for self-affine processes where $qH(q)$ is linear ($H(q)=H$) and fully determined by its index $H$. (Recall that, a transformation is called affine when it scales time and distance by different factors, while a behavior that reproduces itself under affine transformation is called self-affine \cite{FracScaFin}. A time-dependent self-affine function $X(t)$ has fluctuations on different time scales that can be rescaled so that the original signal $X(t)$ is statistically equivalent to its rescaled version $\lambda ^{-H} X(\lambda t)$ for any positive $\lambda$, i.e. $X(t) \sim  \lambda ^{-H} X(\lambda t)$. Brownian motion is self-affine by nature.) In the second case, when $H(q)$ depends on $q$, the process is commonly called multi-scaling (or multi-fractal) and different exponents characterize the scaling of different $q$-moments of the distribution.

\begin{figure}
\begin{center}
\mbox{\epsfig{file=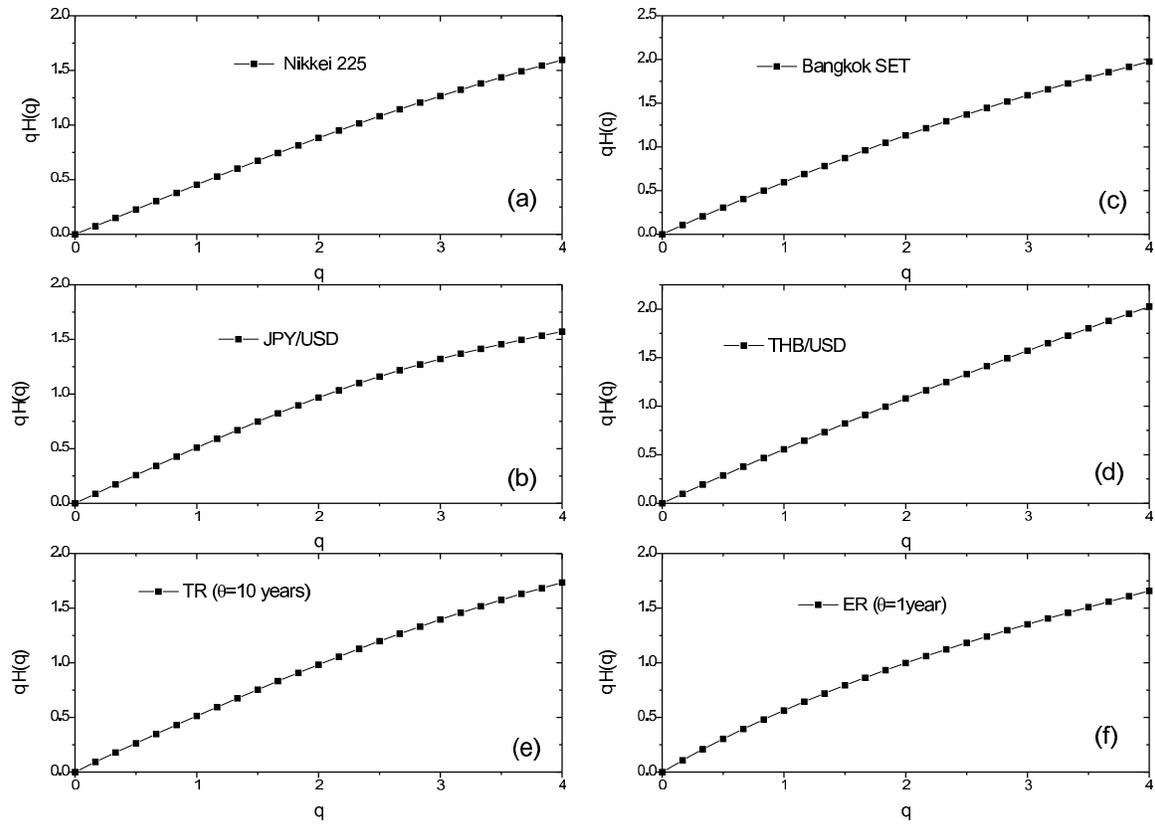,width=6.in,angle=0}}
\end{center}
\caption{The function $q H(q)$ vs. $q$ in the time period from $1997$ to $2001$. (a) JAPAN (Nikkei 225); (b) JAPAN (JPY/USD); (c) Thailand (Bangkok SET); (d) Thailand (THB/USD); (e) Treasury rates having maturity dates $\theta=10$ years; (f) Eurodollar rates having maturity dates $\theta=1$ year. For (f) the time period is $1990$ - $1996$.}
\label{f.HQT}
\end{figure}

For some values of $q$, the exponents are associated with special features. For instance, when $q=1$, $H(1)$ describes the scaling behavior of the absolute values of the increments. The value of this exponent is expected to be closely related to the original Hurst exponent, $H$, that is indeed associated with the scaling of the absolute spread in the increments. The exponent at $q=2$, is associated with the scaling of the autocorrelation function and is related to the power spectrum \cite{PFL.1989-01-01}. A special case is associated with the value of $q=q^*$ at which $q^*H(q^*)=1$. At this value of $q$, the moment $K_{q^*}(\tau)$ scales linearly in $\tau$ \cite{FracScaFin}. Since $qH(q)$ is \emph{in general} a monotonic growing function of $q$, we have that all the moments $H_q(\tau)$ with $q < q^*$  will scale slower than $\tau$, whereas all the moments with $q > q^*$ will scale faster than $\tau$. The point $q^*$ is therefore a threshold value. In this paper we focalize the attention on the case $q=1$ and $2$. Clearly in the uni-fractal case $H(1)=H(2)=H(q^*)$. Their values will be equal to 1/2 for the Brownian motion and they would be equal to $H \not=0.5$ for the fractional Brownian motion. However, for more complex processes, these coefficients do not in general coincide. We thus see that the non-linearity of the empirical function $qH(q)$ is a solid argument against Brownian, fractional Brownian, L\'evy, and fractional L\'evy models, which are all additive models, therefore giving for $qH(q)$ straight lines or portions of straight lines. The curves for $qH(q)$ vs. $q$ are reported in Fig.~\ref{f.HQT} for some of the data. One can observe that, for all these time series, $qH(q)$ is not linear in $q$ but slightly bending below the linear trend. The same behavior holds for the other data. This is a sign of deviation from Brownian, fractional Brownian, L\'evy, and fractional L\'evy models, as already seen in FX rates \cite{UAM.1990-01-01}.

\section{Methodology and Preparation of the Data}
\label{s.metodology}
Let us here recall that the theoretical framework we presented in the previous section is based on the assumption that the process has the scaling property described in Eq.~\ref{Hurstgen}. Moreover, we have implicitly assumed that the scaling properties associated with a given time series stay unchanged across the observation time window $T$. On the other hand, it is well known that financial time series show evidences of variation of their statistical properties with time, and show dependencies on the observation time window $T$. The simplest case which shows such a dependence is the presence of a linear drift ($ \eta t $) added to a stochastic variable ($X(t)= \tilde X(t)+\eta t$) with $\tilde X(t)$ satisfying Eq.~\ref{Hurstgen} and the above mentioned properties of stability within the time window. Clearly, the scaling analysis described in the previous section must be applied to the stochastic component $\tilde X(t)$ of the process. This means that we must subtract the drift $\eta t$ from the variable $X(t)$. To this end one can evaluate $\eta$ from the following relation:
\begin{equation}
\label{avinBeta}
\left< X(t+\tau) - X(t) \right> = \eta \tau \;\;\;.
\end{equation}
Other more complex deviations from the stationary behavior might be present in the financial data that we analyze. In this context, the subtraction of the linear drift can be viewed as a first approximation. 

Our empirical analysis is performed on the daily time series TR, ER, FX and SM (described in Section~\ref{s.data}) which span typically over periods between $1000$ and $3000$ days. In particular, we analyze the time series themselves for the TR and ER, whereas we compute the returns from the logarithmic price $X(t)=\ln(P(t))$ for FX and SM. Moreover, all of these variables are `detrended' by eliminating the linear drift (if there is one) as described in (Eq.~\ref{avinBeta}).

We compute the $q$-order moments $K_q(\tau)$ (defined in Eq.~\ref{multi}) of the `detrended' variables and their logarithms with $\tau$ in the range between $\nu=1$ day and $\tau_{max}$ days. In order to test the robustness of our empirical approach, for each series we analyze the scaling properties varying $\tau_{max}$ between $5$ and $19$ days. We compute the 99$\%$ confidence intervals of all the exponents using different $\tau_{max}$ values\footnote{By using a Matlab routine, namely, normfit that computes parameter estimates and confidence intervals for normal data.}. The resulting exponents computed using different $\tau_{max}$ are stable in their values within a range of $10 \%$. We then verify that the scaling behavior given in Eq.~\ref{Hurstgen} is well followed (see Fig.~\ref{f.nik225_kq_dt}) and we compute the associated generalized Hurst exponent $H(q)$ whose values are given in the following section. We also tested the influence of the detrending (through Eq.~\ref{avinBeta}) calculating the generalized Hurst exponent both for the detrended and the non-detrending time series. The results are in all cases comparable within the standard deviations calculated varying $\tau_{max}$.

\section{Results}
\label{s.res}
\subsection{Computation of the generalized Hurst exponent}
\label{s.comp}
\begin{figure}
\begin{center}
\mbox{\epsfig{file=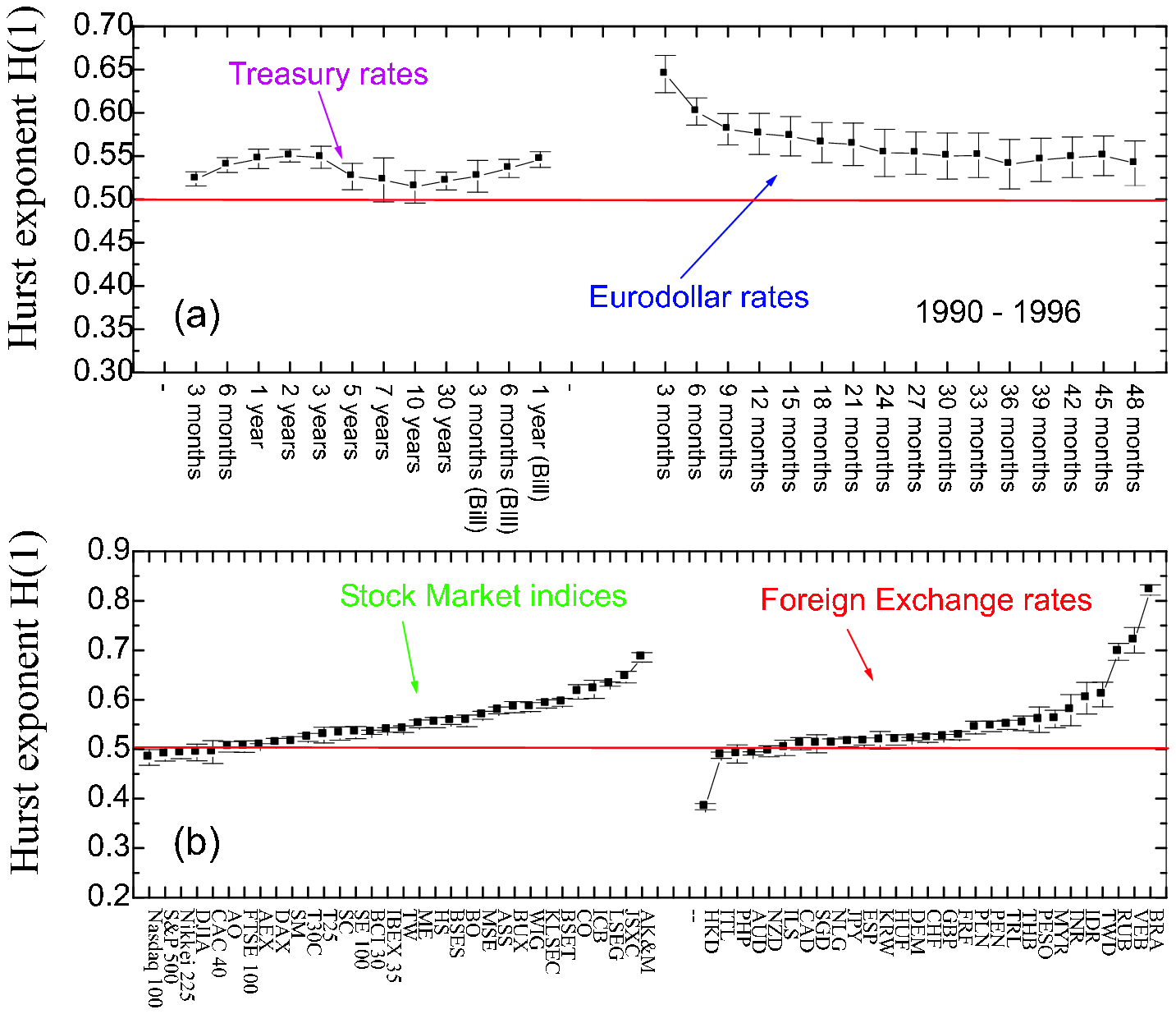,width=6.in,angle=0}}
\end{center}
\caption{(a) The Hurst exponent $H(1)$ for the Treasury and Eurodollar rates time series in the period from $1990$ to $1996$; (On the $x$-axis the corresponding maturities dates are reported.) (b) The Hurst exponent $H(1)$ for the Stock Market indices and Foreign Exchange rates in the time period reported in Tabs. ~\ref{t.1} and ~\ref{t.2}. (On the $x$-axis the corresponding data-sets are reported.)}
\label{f.H1log}
\end{figure}

\begin{figure}
\begin{center}
\mbox{\epsfig{file=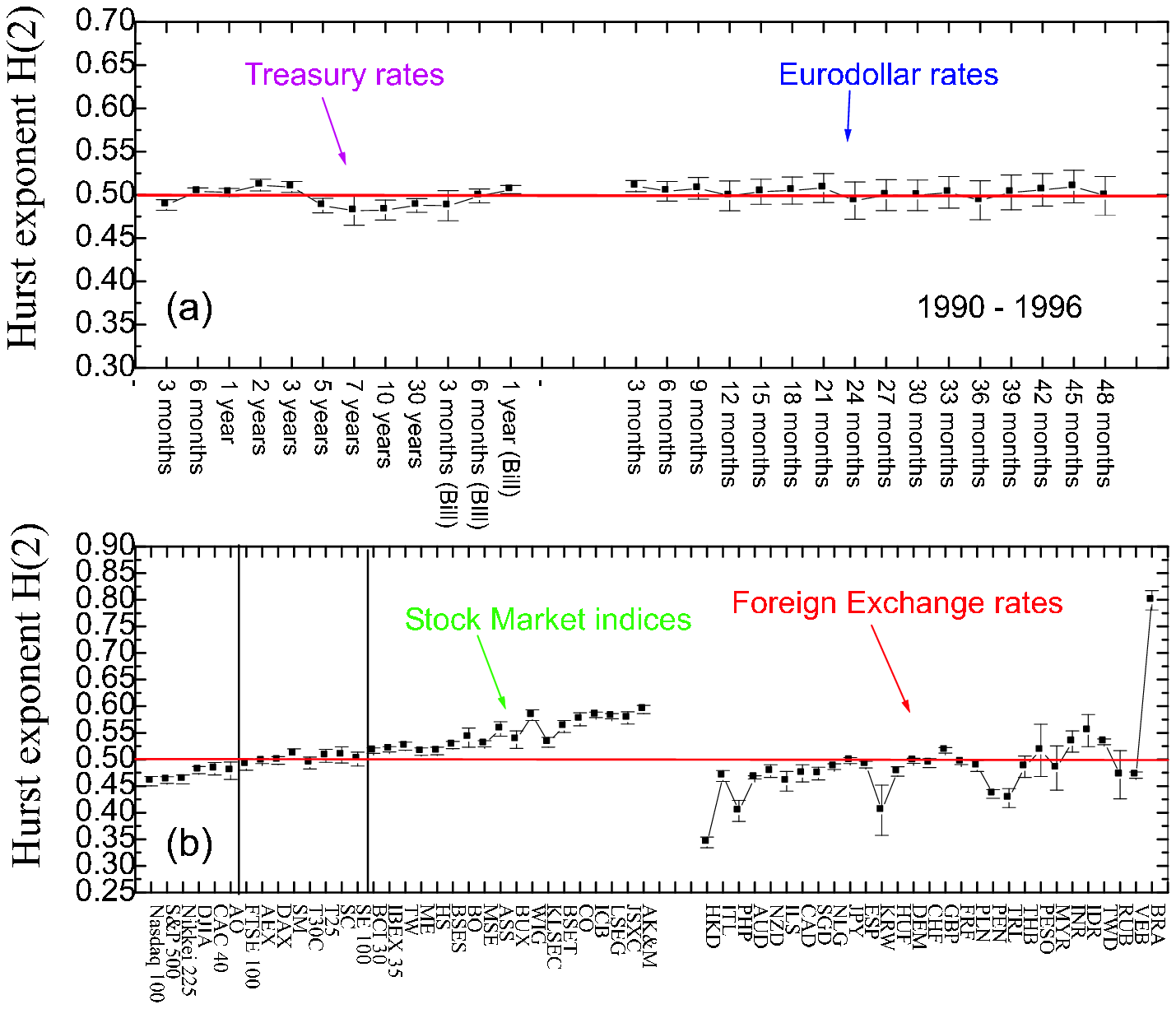,width=6.in,angle=0}}
\end{center}
\caption{(a) The Hurst exponent $H(2)$ for the Treasury and Eurodollar rates time series in the period from $1990$ to $1996$; (On the $x$-axis the corresponding maturities dates are reported.) (b) The Hurst exponent $H(2)$ for the Stock Market indices and Foreign Exchange rates in the time period reported in Tabs. ~\ref{t.1} and ~\ref{t.2}. (On the $x$-axis the corresponding data-sets are reported.)}
\label{f.H2log}
\end{figure}

In this section we report and discuss the results for the scaling exponents $H(q)$ computed for $q=1$ and $q=2$. These exponents $H(1)$ and $H(2)$ for all the assets and different markets (presented in Section~\ref{s.data}) are reported in Figs.~\ref{f.H1log}~and~\ref{f.H2log} respectively. Figures~\ref{f.H1log}~(a)~and~\ref{f.H2log}~(a) refer to the Treasury and Eurodollar rates in the time period from $1990$ to $1996$. Whereas Figures~\ref{f.H1log}~(b)~and~\ref{f.H2log}~(b) are relative to the Stock Market indices and Foreign Exchange rates in the time period reported in Tables~\ref{t.1}~and~\ref{t.2}. The data points are the average values of $H(1)$ and $H(2)$ computed from a set of values corresponding to different $\tau_{max}$ (between $5$ and $19$ days) and the error bars are their standard deviations. The generalized Hurst exponents are computed through a linear least squares fitting. We have computed the standard deviations for the two linear fit coefficients and the correlation coefficient. It results that the standard deviations from the linear fitting are below or equal to the reported standard deviations values computed varying $\tau_{max}$. The correlation coefficient is never lower that $0.99$. 

Let us first notice that, for fixed income instruments (Figs.~\ref{f.H1log}(a)~and~\ref{f.H2log}(a)), $H(2)$ is close to $0.5$ while $H(1)$ is rather systematically above $0.5$ (with the $3$ months Eurodollar rate that shows a more pronounced deviation because it is directly influenced by the actions of central banks). On the other hand, as far as Stock markets are concerned, we find that the generalized Hurst exponents $H(1)$, $H(2)$ show remarkable differences between developed and emerging markets. In particular, the values of $H(1)$, plotted in Fig.~\ref{f.H1log}(b), present a differentiation across $0.5$ with high values of $H(1)$ associated with the emerging markets and low values of $H(1)$ associated with developed ones. In Fig.~\ref{f.H1log}(b) the ordering of the stock markets from left to right is chosen in ascending order of $H(1)$. One can see that such a ordering corresponds very much to the order one would intuitively give in terms of maturity of the markets. Moreover, we can see from Fig.~\ref{f.H2log}(b) that the different assets can be classified into three different categories:
\begin {enumerate}
\item
First those that have an exponent $H(2)>0.5$ which includes all indices of the emerging markets and the BCI $30$ (Italy), IBEX $35$ (Spain) and the Hang Seng (Hong Kong).
\item
A second category concerns the data exhibiting $H(2) \sim  0.5$ (within the error bars). This category includes: FTSE $100$ (UK), $AEX$ (Netherlands), $DAX$ (Germany), Swiss Market (Switzerland), Top $30$ Capital (New Zealand), Tel Aviv $25$ (Israel), Seoul Composite (South Korea) and Toronto SE $100$ (Canada).
\item
A third category is associated with $H(2)<0.5$ and includes the following data: Nasdaq $100$ (US), $S\&P 500$ (US), Nikkei $225$ (Japan), Dow Jones Industrial Average (US), CAC $40$ (France) and All Ordinaries (Australia).
\end{enumerate}
We find therefore that all the emerging markets have $H(2) \geq  0.5$ whereas all the well developed have $H(2) \leq 0.5$. This simple classification is not achieved by other means. One could, for instance use the Sharpe Ratio \cite{WFS.1994-01-01} we have tried it but it does not achieve such a clear cut categorization. This ratio requires a benchmark risk free return that is not always available for emerging markets. We have tried the classification obtained from a simple ratio of the average returns of their standard deviations but the ordering is not conclusive.

We find that the Foreign Exchange rates show $H(1)>0.5$ quite systematically. This is consistent with previous results computed with high frequency data \cite{UAM.1990-01-01}, although the values here are slightly lower. An exception with pronounced $H(1)<0.5$ is the HKD/USD (Hong Kong) (Fig.~\ref{f.H1log}~(b)). This FX rate is, or has been, at one point pegged to the USD, that is why its exponent differs from the others. Whereas in the class $H(1) \sim  0.5$ we have: $ITL/USD$ (Italy), $PHP/USD$ (Philippines), $AUD/USD$ (Australia), NZD/USD (New Zealand), ILS/USD (Israel), $CAD/USD$ (Canada), $SGD/USD$ (Singapore), $NLG/USD$ (Netherlands) and $JPY/USD$ (Japan). On the other hand, the values of $H(2)$ (Fig.~\ref{f.H2log}~(b)) show a much larger tendency to be $<0.5$ with some stronger deviations such as: $HKD/USD$ (Hong Kong), $PHP/USD$ (Philippines), $KRW/USD$(South Korea), $PEN/USD$ (Peru) and $TRL/USD$ (Turkey). Whereas values of $H(2)>0.5$ are found in: $GBP/USD$ (United Kingdom), $PESO/USD$ (Mexico), $INR/USD$ (India), $IDR/USD$ (Indonesia), $TWD/USD$ (Taiwan) and $BRA/USD$ (Brazil).

\subsection{Checking the Temporal and Numerical Stability of the Results}
\label{s.stab}
\begin{figure}
\begin{center}
\mbox{\epsfig{file=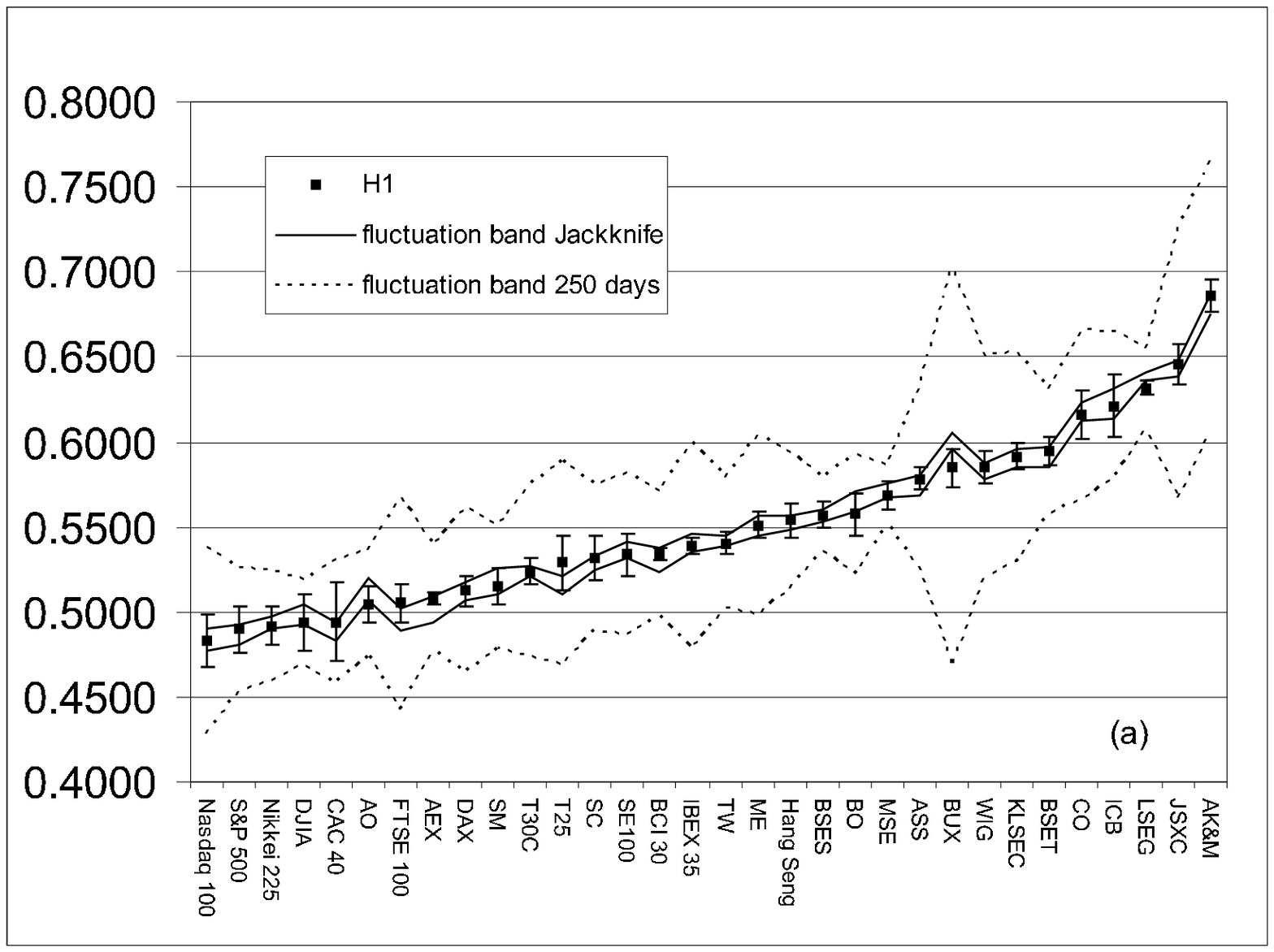,width=4.in,angle=0}}
\mbox{\epsfig{file=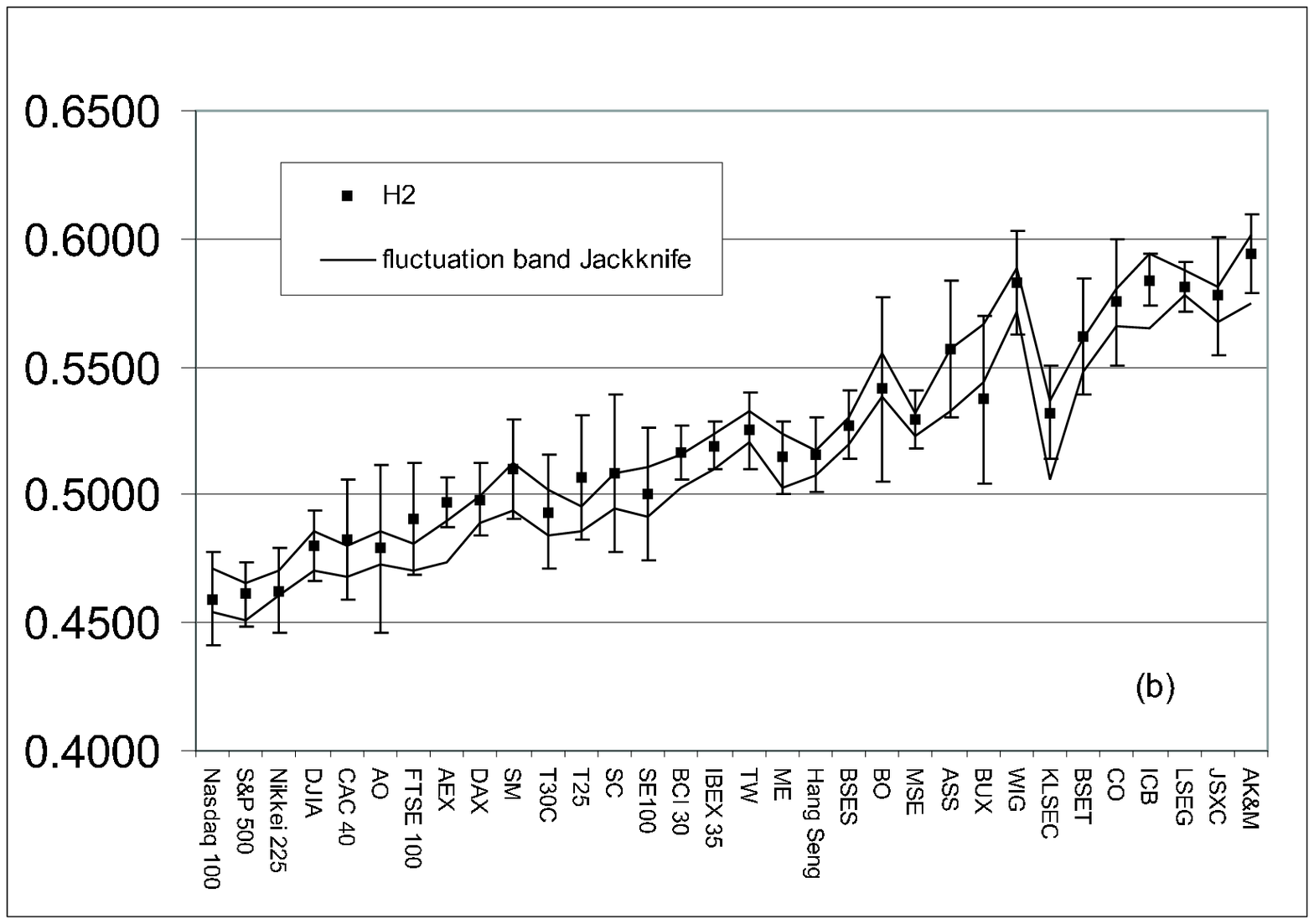,width=4.in,angle=0}}
\end{center}
\caption{(a) The generalized Hurst exponent $H(1)$ for the Stock Market indices in the whole time period (see Tab. ~\ref{t.2}) with its variation (black lines) obtained by using the Jackknife method and its variation (dashed lines) when time periods of $250$ days are considered; (b) The generalized Hurst exponent $H(2)$ for the Stock Market indices in the whole time period (see Tab. ~\ref{t.2}) with its variation (black lines) obtained by using the Jackknife method. The square points are the average values of $H(1)$ and $H(2)$ computed from a set of values corresponding to different $\tau_{max}$. The error bars are their standard deviations.}
\label{f.Fig6}
\end{figure}

In order to check the temporal stability of the results, these analysis are performed also over different time periods and the values of the exponents $H(1)$ and $H(2)$ are reported in Table~\ref{t.Htimeperiod9701} for the time period from $1997$ to $2001$ for Foreign Exchange rates, Stock Market indices and Treasury rates. These results should be compared with those obtained on the whole time period (shown in Table~\ref{t.2}) and to time periods of $250$ days. Moreover, we tested the numerical robustness of our results by using the Jackknife method \cite{HRK.1989-01-01} which consists of taking out randomly $1/10$ of the sample and iterates the procedure $10$ times (every time taking out data which were not taken out in the previous runs). On one hand, we observe (see Fig.~\ref{f.Fig6}) that the generalized Hurst exponents computed on these Jackknife-reduced time series are very close to those computed on the entire series with deviations inside the errors estimated by varying $\tau_{max}$ (as described in section~\ref{s.metodology}). This indicates a strong numerical stability. On the other hand, the analysis on sub-periods of $250$ days shows fluctuations that are larger than the previous estimated errors (and larger than the variations with the Jackknife method) indicating therefore that there are significant changes in the market behaviors over different time periods (Fig.~\ref{f.Fig6}~(a)). This phenomenon was also detected in \cite{HFIntro} when studying Exchange rates that were part of the European Monetary System. It seems that $H(1)$ is particularly sensitive to institutional changes in the market. The scaling exponents cannot be assumed to be constant over time if a market is experiencing major institutional changes. Nevertheless, well developed markets have values of $H(2)$ that are on average smaller than the emerging ones and the weakest markets have oscillation bands that stay above $0.5$ whereas the strongest have oscillation bands that contain $0.5$.

Numerical stability gives us confidence in our method for determining the exponent and temporal variability is a sign that the exponents are sensitive to institutional changes in the market reinforcing our idea to use them as indicators of the maturity of the market.

\section{Scaling exponents in the frequency domain}
\label{s.frescaling}
\subsection{Spectral analysis}
\label{s.spectral}
\begin{figure}
\begin{center}
\mbox{\epsfig{file=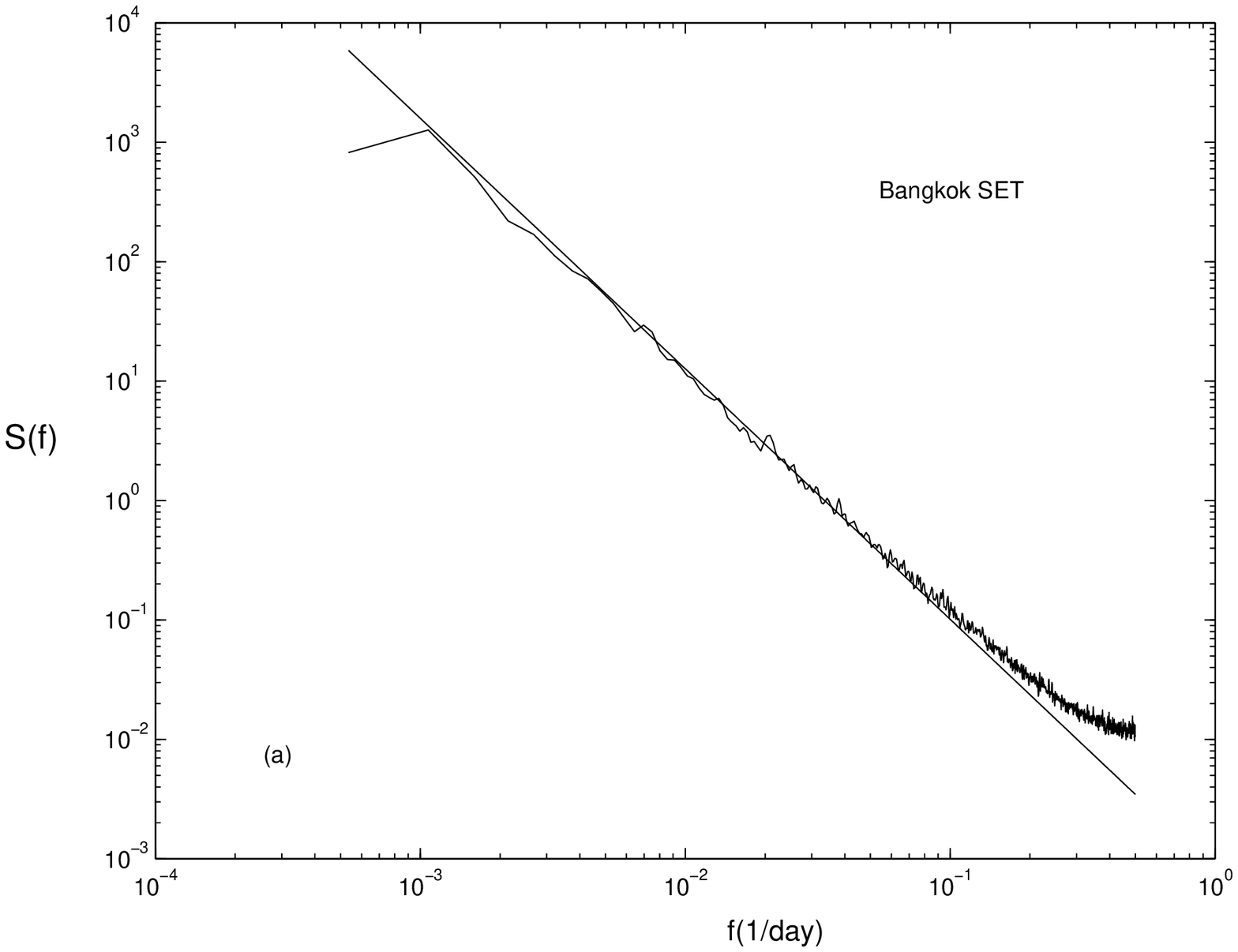,width=3.in,angle=0}}
\mbox{\epsfig{file=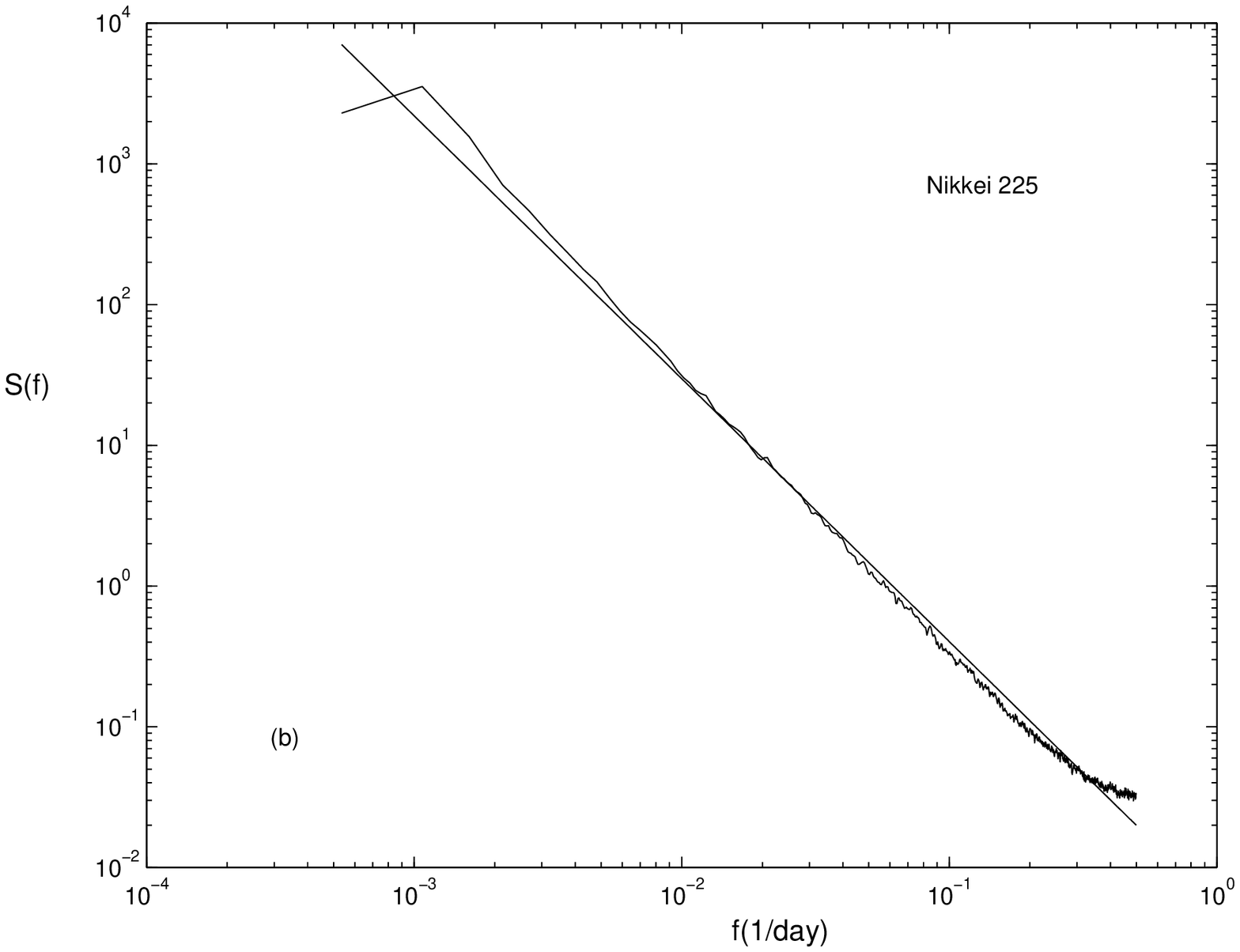,width=3.in,angle=0}}
\end{center}
\caption{The power spectra of the Stock Market indices compared with the behaviour of $f^{-2H(2)-1}$ (straight lines in log-log scale) computed using the Hurst exponents values in the time period $1997$-$2001$; (a) Thailand (Bangkok SET) and (b) JAPAN (Nikkei 225).
The line is the prediction from the generalized Hurst exponent $H(2)$ (Eq. ~\ref{www}).}

\label{f.SpettroSM}
\end{figure}

In order to empirically investigate the statistical properties of the time series in the frequency domain we perform a spectral analysis computing the power spectral density (PSD) \cite{SMK.1981-01-01} by using the periodogram approach, that is currently one of the most popular and computationally efficient PSD estimator. This is a sensitive way to estimate the limits of the scaling regime of the data increments. The results for some SM data in the time periods $1997$-$2001$ are shown in Fig.~\ref{f.SpettroSM}. For SM we compute the power spectra of the logarithm of these time series. As one can see the power spectra show clear power law behaviors: $S(f) \sim f^{-\beta }$. This behavior holds for all the other data.

\begin{figure}
\begin{center}
\mbox{\epsfig{file=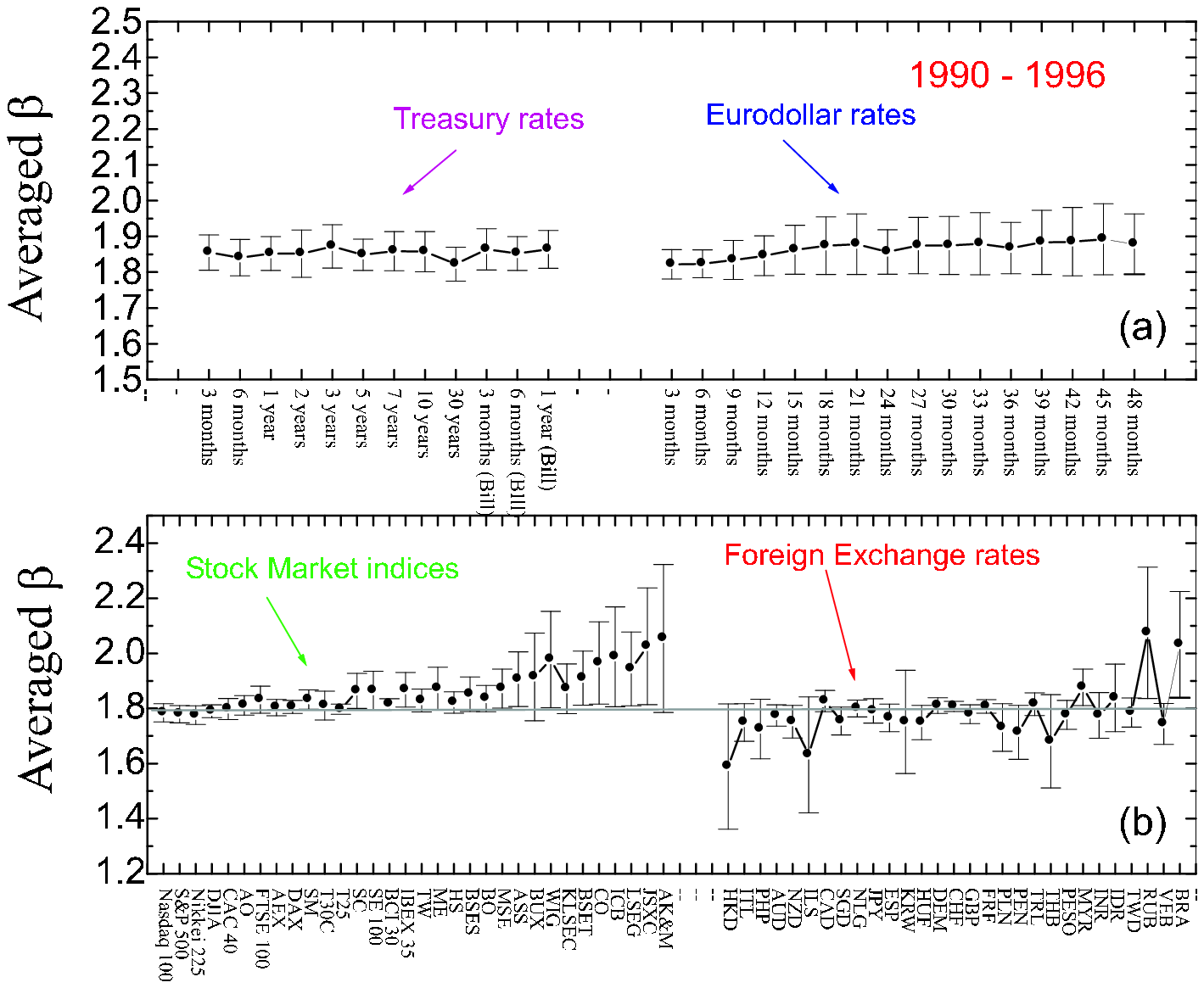,width=6.in,angle=0}}
\end{center}
\caption{(a) The averaged $\beta$ values computed from the power spectra (mean square regression) of the Treasury and Eurodollar rates time series in the period from $1990$ to $1996$; (On the $x$-axis the corresponding maturities dates are reported.) (b) The averaged $\beta$ values computed from the power spectra of the Stock Market indices and Foreign Exchange rates in the time period reported in Tabs. ~\ref{t.1} and ~\ref{t.2}. The horizontal gray line corresponds to the value of $\beta$ obtained from the simulated random walks reported in Table ~\ref{t.HbetaGaussian}. (On the $x$-axis the corresponding data-sets are reported.)}
\label{f.beta}
\end{figure}

The non-stationary features have been investigated by varying the window-size on which the spectrum is calculated from $100$ days up to the entire size of the time series. The power spectra coefficients $\beta$ are calculated through a mean square regression in log-log scale. The values reported in Fig.~\ref{f.beta} are the average of the evaluated $\beta$ over different windows and the error bars are their standard deviations. Fig.~\ref{f.beta}(a) refers to a time period between $1990$ to $1996$ whereas the Stock Market indices and Foreign Exchange rates (Fig.~\ref{f.beta}(b)) are analyzed over the time periods reported in Tables~\ref{t.1}~and~\ref{t.2}. Moreover, the averaged $\beta$ values in a different time period, namely from $1997$ to $2001$ are reported in Table~\ref{t.betatimeperiod9701} for Foreign Exchange rates, Stock Market indices and Treasury rates. These values differ from the spectral density exponent expected for a pure Brownian motion ($\beta=2$). However, we will shown in Section~\ref{s.monte} that this method is biased and we indeed found power spectra exponents around $1.8$ for random walks using three different random numbers generators.

It must be noted that, the power spectrum is only a second order statistic and its slope is not enough to validate a particular scaling model: it gives only partial information about the statistics of the process.

\subsection{Scaling Spectral Density and Hurst Exponent}
\label{s.beta}
For financial time series, as well as for many other stochastic processes, the spectral density $S(f)$ is empirically found to scale with the frequency $f$ as a power law: $S(f) \propto f^{-\beta}$ as already stated in the previous section. Here we use a simple argument to show how this scaling in the frequency domain should be related to the scaling in the time-domain. Indeed, it is known that the spectrum $S(f)$ of the signal $X(t)$ can be conveniently calculated from the Fourier transform of the autocorrelation function (Wiener-Khinchin theorem). On the other hand, the autocorrelation function of $X(t)$ is proportional to the second moment of the distribution of the increments which, from Eq.~\ref{Hurstgen}, is supposed to scale as $K_2 \sim \tau^{2H(2)}$. But, the components of the Fourier transform of a function which behaves in the time-domain as $\tau^{\alpha}$ are proportional to $f^{-\alpha-1}$ in the frequency-domain. Therefore, we have that the power spectrum of a signal that scales as Eq.~\ref{Hurstgen} must behave as:
\begin{equation}
\label{www}
S(f) \propto f^{-2H(2)-1} \;\;\;.
\end{equation}
Consequently, the slope $\beta$ of the power spectrum is related to the generalized Hurst exponent for $q=2$ through: $\beta=1+2H(2)$. Note that Eq.~\ref{www} is obtained only assuming that the signal $X(t)$ has a scaling behavior in accordance to Eq.~\ref{Hurstgen} without making any hypothesis on the kind of underlying mechanism that might lead to such a scaling behavior.

We here compare the behavior of the power spectra $S(f)$ with the function $f^{-2H(2)-1}$ which - according to Eq.~\ref{www} - is the scaling behavior expected in the frequencies domain for a time series which scales in time with a generalized Hurst exponent $H(2)$. We performed such a comparison for all the financial data and we report in Fig.~\ref{f.SpettroSM} those for Stock Market indices for Thailand and JAPAN (in the time period $1997$-$2001$). As one can see the agreement between the power spectra behavior and the prediction from the generalized Hurst analysis is very satisfactory. This result holds also for all the other data. Note that the values of $2 H(2)+1$ do not in general coincide with the values of the power spectral exponents evaluated by means of the mean square regression. The method through the generalized Hurst exponent appears to be more powerful in catching the scaling behavior even in the frequency domain.

\section{Monte Carlo Test of the Method}
\label{s.monte}
In the literature, the scaling analysis has been criticized for being biased. In order to test that our method is not biased we estimate the generalized Hurst exponents for simulated random walks. We produce synthetic time series by using three different random number generators. We perform $100$ simulations of random walks with the same number of data points as in our samples ($991$ and $3118$) and estimate the generalized Hurst exponents $H(1)$ and $H(2)$ and the power spectra exponents $\beta$. The results are reported in Table~\ref{t.HbetaGaussian}.

In all the cases, $H(1)$ and $H(2)$ have values of $0.5$ within the errors. Only when we consider uniformly distributed random numbers in the interval ($0$,$1$) (Rand which uses a lagged Fibonacci generator combined with a shift register random integer generator, based on the work of Marsaglia \cite{GMA.1994-01-01}.) we obtain for $H(1)$ of $0.47 \pm 0.01$, but also in this case $H(2)$ is 0.5 within the errors. On one hand, this shows that our method is powerful and robust and is not biased as other methods are. On the other hand, the estimations of $\beta$ from the power spectrum have values around $1.8$ (instead of $2$), showing therefore that this other method is affected by a certain bias.

\section{Conclusion}
\label{s.conclusion}
By applying the same methodology to a wide variety of markets and instruments (89 in total), this study confirms that empirical scaling behaviors are rather universal across financial markets. By analyzing the scaling properties of the $q$-order moments (Eq.~\ref{multi}) we show that the generalized Hurst exponent $H(q)$ (Eq.~\ref{Hurstgen}) is a powerful tool to characterize and differentiate the structure of such scaling properties. Our study also confirms that $qH(q)$ exhibits a non-linear dependence on $q$ which is a clear signature of deviations from pure Brownian motion and other additive or uni-scaling models.

The novelty of this work resides in the empirical analysis across a wide variety of stock indices that shows the sensitivity of the exponent $H(2)$ to the degree of development of the market. At one end of the spectrum, we find: the Nasdaq $100$ (US), the $S\&P 500$ (US), the Nikkei $225$ (Japan), the Dow Jones Industrial Average (US), the CAC $40$ (France) and the All Ordinaries index (Australia); all with $H(2)<0.5$. Whereas, at the other end, we find the Russian AK\&M, the Indonesian JSXC, the Peruvian LSEG, etc. (Fig.~\ref{f.H2log}(b)); all with $H(2)>0.5$. Moreover, we observe emerging structures in the scaling behaviors of interest rates and exchange rates that are related to specific conditions of the markets. For example, a strong deviation of the scaling exponent for the 3 months maturity, which is strongly influenced by the central bank decisions. This sensitivity of the scaling exponents to the market conditions provides a new and simple way of empirically characterizing the development of financial markets. Other methods usually used for controlling risk, like standard deviation or Sharpe Ratio are not able to provide such a good classification. 
The robustness of the present empirical approach is tested in several ways: by first comparing theoretical exponents with the results of Monte Carlo simulations using three distinct random generators, second by varying the maximum time-step ($\tau_{max}$) in the analysis, third by applying the Jackknife method to produce several samples, fourth by varying the time-window sizes to analyze the temporal stability and fifth by computing results for detrended and non-detrending time series. We verify that the observed differentiation among different degrees of market development is clearly emerging well above the numerical fluctuations. Finally, from the comparison between the empirical power spectra and the prediction from the scaling analysis (Eq.~\ref{www}, Fig.~\ref{f.SpettroSM}) we show that the method through the generalized Hurst exponent describes well the scaling behavior even in the frequency domain.

\section*{Acknowledgments}
T. Di Matteo wishes to thank Sandro Pace for fruitful discussions
and support. M. Dacorogna benefited from discussions with the
participants to the CeNDEF workshop in Leiden, June 2002.

\newpage
%\bibliographystyle{obib}
%\bibliography{books,journals}

\newpage

\vspace*{3cm}

\vspace*{2cm}

%\newpage

\vspace*{2cm}

\newpage

\vspace*{2cm}

\newpage

\vspace*{2cm}

\newpage

\newpage

\vspace*{1cm}

\newpage

\vspace*{3cm}

\newpage
\begin{table}
\caption{Foreign Exchange rates (FX/USD).}
\label{t.1}
\begin{tabular}{cccccc}
$ Country $ & $FX$ & Time period & $ Country $ & $FX$ & Time period \\
\hline
Hong Kong & HKD & 1990-2001 & United Kingdom & GBP & 1990-2001 \\
Italy & ITL & 1993-2001 & France & FRF & 1993-2001 \\
Philippines & PHP & 1991-2001 & Poland & PLN & 1993-2001 \\

Australia & AUD & 1990-2001 & Peru & PEN & 1993-2001 \\
New Zealand & NZD & 1990-2001 & Turkey & TRL & 1992-2001 \\
Israel & ILS & 1990-2001 & Thailand & THB & 1990-2001 \\
Canada & CAD & 1993-2001 & Mexico & PESO & 1993-2001 \\
Singapore & SGD & 1990-2001 & Malaysia & MYR & 1990-2001 \\
Netherlands & NLG & 1993-2001 & India & INR & 1990-2001 \\
Japan & JPY & 1990-2001 & Indonesia & IDR & 1991-2001 \\
Spain & ESP & 1990-2001 & Taiwan & TWD & 1990-2001 \\
South Korea & KRW & 1990-2001 & Russia & RUB & 1993-2001 \\
Hungary & HUF & 1993-2001 & Venezuela & VEB & 1993-2001 \\
Germany & DEM & 1990-2001 & Brazil & BRA & 1993-2001 \\
Switzerland & CHF & 1993-2001 & \\
\end{tabular}
\end{table}

%\newpage
\begin{table}
\caption{Stock Market indices (SM).}
\label{t.2}
\begin{tabular}{ccc}
$ Country $ & $SM$ & Time period \\
\hline
United States & Nasdaq 100 & 1990-2001 \\
United States & S\&P 500 & 1987-2001 \\
Japan & Nikkei 225 & 1990-2001 \\
United States & Dow Jones Industrial Average (DJIA) & 1990-2001 \\
France & CAC 40 & 1993-2001 \\
Australia & All Ordinaries (AO) & 1992-2001 \\
United Kingdom & FTSE 100 & 1990-2001 \\
Netherlands & AEX & 1993-2001 \\
Germany & DAX & 1990-2001 \\
Switzerland & Swiss Market (SM) & 1993-2001 \\

New Zealand & Top 30 Capital (T30C) & 1992-2001 \\
Israel & Telaviv 25 (T25) & 1992-2001 \\
South Korea & Seoul Composite (SC) & 1990-2001 \\
Canada & Toronto SE 100 (SE 100) & 1993-2001 \\
Italy & BCI 30 & 1993-2001 \\
Spain & IBEX 35 & 1990-2001 \\
Taiwan & Taiwan Weighted (TW) & 1990-2001 \\
\end{tabular}
\end{table}

\begin{table}

%\caption{Table 2}
\begin{tabular}{ccc}
Table $2$ (continued)\\
$ Country $ & $SM$ & Time period \\
\hline
Argentina & Merval (ME) & 1993-2001 \\
Hong Kong & Hang Seng (HS) & 1990-2001 \\
India & Bombay SE Sensex (BSES) & 1990-2001 \\
Brazil & Bovespa (BO) & 1993-2001 \\
Mexico & Mexico SE (MSE) & 1993-2001 \\
Singapore & All Singapore Shared (ASS) & 1990-2001 \\
Hungary & Budapest BUX (BUX) & 1993-2001 \\
Poland & Wig (WIG) & 1991-2001 \\
Malaysia & KLSE Composite (KLSEC) & 1990-2001 \\
Thailand & Bangkok SET (BSET) & 1990-2001 \\
Philippines & Composite (CO) & 1990-2001 \\
Venezuela & Indice de Cap. Bursatil (ICB) & 1993-2001 \\
Peru & Lima SE General (LSEG) & 1993-2001 \\
Indonesia & JSX Composite (JSXC) & 1990-2001 \\
Russia & AK\&M Composite (AK\&M) & 1993-2001 \\
\end{tabular}
\end{table}

\newpage
\begin{table}

\caption{Treasury rates ($TR_i({\theta})$).}
\label{t.3}
\begin{tabular}{cccccccccccc}
$ i $ & $\theta$ & $ i $ & $\theta$ \\
\hline
1 & 3 months & 7 & 7 years  \\
2 & 6 months & 8 & 10 years \\
3 & 1 year & 9 & 30 years   \\
4 & 2 years & 10 & 3 months (Bill)   \\
5 & 3 years & 11 & 6 months (Bill) \\
6 & 5 years & 12 & 1 year (Bill) \\
\end{tabular}
\end{table}

\newpage
\begin{table}
\caption{Eurodollar rates ($ER_i({\theta})$).}
\label{t.4}
\begin{tabular}{cccccccccccc}
$ i $ & $\theta$ & $ i $ & $\theta$ \\
\hline
1 & 3 months & 9 & 27 months \\
2 & 6 months & 10 & 30 months   \\
3 & 9 months & 11 & 33 months \\
4 & 12 months & 12 & 36 months \\
5 & 15 months & 13 & 39 months \\
6 & 18 months & 14 & 42 months \\
7 & 21 months & 15 & 45 months \\
8 & 24 months & 16 & 48 months \\
\end{tabular}
\end{table}

\newpage

\begin{table}
\caption{Hurst exponents $H(1)$ and $H(2)$ and averaged $\beta$ values computed for random walks simulated by using three different random numbers generators: 1) Randn=Normally distributed random numbers with mean $0$ and variance $1$; 2) Rand=Uniformly distributed random numbers in the interval ($0,1$) and 3) Normrnd=Random numbers from the normal distribution with mean $0$ and standard deviation $1$. These are average values on $100$ simulations of random walks with $991$ and $3118$ numbers of data points.}
\label{t.HbetaGaussian}
\begin{tabular}{cccccccccccc}
$ N $ & $ H(1) $ & $ H(2) $ & $\beta$ \\
\hline
1) Randn \\
$991$ & $ 0.50 \pm 0.01 $ & $0.50 \pm 0.01 $ & $ 1.8 \pm 0.1 $ \\
$3118$ & $ 0.50 \pm 0.01 $ & $0.50 \pm 0.01 $ & $ 1.80 \pm 0.03 $ \\
2) Rand \\
$991$ & $ 0.47 \pm 0.01 $ & $ 0.49 \pm 0.01 $ & $ 1.8 \pm 0.1 $ \\
$3118$ & $ 0.47 \pm 0.01 $ & $ 0.50 \pm 0.01 $ & $ 1.80 \pm 0.03 $ \\
3) Normrnd \\
$991$ & $ 0.49 \pm 0.01 $ & $0.49 \pm 0.01 $ & $ 1.8 \pm 0.1 $ \\
$3118$ & $ 0.50 \pm 0.01 $ & $0.50 \pm 0.01 $ & $ 1.80 \pm 0.03 $ \\
\end{tabular}
\end{table}

\newpage

\begin{table}
\caption{Hurst exponents $H(1)$ and $H(2)$ for Foreign Exchange rates, Stock Market indices and Treasury rates in the time period from $1997$ to $2001$.}
\label{t.Htimeperiod9701}
\begin{tabular}{cccccccccccc}
$ Data $ & $ H(1) $ & $ H(2)$ & $ Data $ & $ H(1) $ & $ H(2)$ \\
\hline

Foreign Exchange rates \\
HKD & $0.41 \pm 0.01$ & $0.34 \pm 0.01$ & GBP & $0.50 \pm 0.02$ & $0.48 \pm 0.02$    \\
ITL & $0.51 \pm 0.01$ & $0.51 \pm 0.01$  & FRF & $0.51 \pm 0.01$ & $0.51 \pm 0.01$   \\

PHP & $0.52 \pm 0.01$ & $0.43 \pm 0.02$  & PLN & $0.54 \pm 0.01$ & $0.50 \pm 0.01$  \\
AUD & $0.52 \pm 0.01$ & $0.502 \pm 0.002$ & PEN & $0.52 \pm 0.01$ & $0.41 \pm 0.03$ \\
NZD & $0.49 \pm 0.01$ & $0.48 \pm 0.01$ &  TRL & $0.56 \pm 0.01$ & $0.44 \pm 0.04$    \\
ILS & $0.48 \pm 0.02$ & $0.47 \pm 0.02$ & THB & $0.53 \pm 0.01$ & $0.50 \pm 0.02$   \\
CAD & $0.51 \pm 0.01$ & $0.48 \pm 0.01$  & PESO & $0.53 \pm 0.01$ & $0.50 \pm 0.01$  \\
SGD & $0.50 \pm 0.01$ & $0.47 \pm 0.03$ & MYR & $0.51 \pm 0.03$ & $0.45 \pm 0.05$    \\
NLG & $0.51 \pm 0.01$ & $0.51 \pm 0.01$ &  INR & $0.58 \pm 0.02$ & $0.53 \pm 0.01$   \\
JPY & $0.50 \pm 0.01$ & $0.49 \pm 0.01$ &  IDR & $0.56 \pm 0.03$ & $0.53 \pm 0.03$  \\
ESP & $0.50 \pm 0.01$ & $0.49 \pm 0.01$ & TWD & $0.58 \pm 0.01$ & $0.51 \pm 0.01$  \\
KRW & $0.50 \pm 0.03$ & $0.39 \pm 0.06$ & RUB & $0.64 \pm 0.02$ & $0.47 \pm 0.03$  \\
HUF & $0.52 \pm 0.01$ & $0.52 \pm 0.01$   & VEB & $0.54 \pm 0.04$ & $0.49 \pm 0.02$    \\
DEM & $0.51 \pm 0.01$ & $0.51 \pm 0.01$  & BRA & $0.59 \pm 0.02$ & $0.60 \pm 0.01$  \\
CHF & $0.51 \pm 0.01$ & $0.50 \pm 0.01$   \\

\end{tabular}
\end{table}

\begin{table}
\begin{tabular}{cccccccccccc}
Table $6$ (continued)\\
$ Data $ & $ H(1) $ & $ H(2)$ & $ Data $ & $ H(1) $ & $ H(2)$ \\
\hline
Stock Market indices\\
Nasdaq 100 & $0.47\pm 0.01$ & $0.45\pm 0.01$  & TW & $0.53 \pm 0.01$ & $0.51 \pm 0.01$   \\
S\&P 500 & $0.47 \pm 0.02$ & $0.44 \pm 0.01$  & ME & $0.57 \pm 0.01$ & $0.53 \pm 0.01$  \\
Nikkei 225 & $0.46 \pm 0.01$ & $0.43 \pm 0.01$ & HS & $0.53 \pm 0.01$ & $0.49 \pm 0.01$   \\
DJIA & $0.49 \pm 0.01$ & $0.464 \pm 0.004$ & BSES & $0.54 \pm 0.01$ & $0.52 \pm 0.01$  \\
CAC 40 & $0.47 \pm 0.02$ & $0.46 \pm 0.02$  & BO & $0.51 \pm 0.01$ & $0.48 \pm 0.01$  \\
AO & $0.49 \pm 0.02$ & $0.46 \pm 0.03$ & MSE & $0.57 \pm 0.01$ & $0.52 \pm 0.01$ \\
FTSE 100 & $0.46 \pm 0.02$ & $0.44 \pm 0.01$ & ASS & $0.57 \pm 0.01$ & $0.54 \pm 0.02$ \\
AEX & $0.49 \pm 0.01$ & $0.47 \pm 0.02$ & BUX & $0.52 \pm 0.01$ & $0.49 \pm 0.01$ \\
DAX & $0.50 \pm 0.01$ & $0.47 \pm 0.01$ & WIG & $0.49 \pm 0.01$ & $0.44 \pm 0.01$  \\
SM & $0.50 \pm 0.02$ & $0.48 \pm 0.02$& KLSEC & $0.60 \pm 0.01$ & $0.51 \pm 0.02$  \\
T30C & $0.49 \pm 0.01$ & $0.46 \pm 0.01$  & BSET & $0.59 \pm 0.01$ & $0.55 \pm 0.01$\\
T25 & $0.53 \pm 0.01$ & $0.51 \pm 0.01$& CO & $0.59 \pm 0.01$ & $0.54\pm 0.01$    \\
SC & $0.53 \pm 0.01$ & $0.51 \pm 0.01$  &   ICB & $0.61 \pm 0.02$ & $0.55 \pm 0.02$      \\
SE 100 & $0.51 \pm 0.01$ & $0.48 \pm 0.01$& LSEG & $0.61 \pm 0.01$ & $0.58 \pm 0.01$  \\
BCI 30 & $0.52 \pm 0.01$ & $0.48 \pm 0.01$& JSXC & $0.57 \pm 0.02$ & $0.53 \pm 0.02$      \\
IBEX 35 & $0.50 \pm 0.01$ & $0.48 \pm 0.01$& AK\&M & $0.65 \pm 0.03$ & $0.51 \pm 0.01$     \\
\end{tabular}
\end{table}

\begin{table}
\begin{tabular}{cccccccccccc}
Table $6$ (continued)\\
$ Data $ & $ H(1) $ & $ H(2)$ \\
\hline
Treasury rates\\
$TR_1$ & $0.48 \pm 0.01$ & $0.44 \pm 0.02$     \\
$TR_2$ & $0.55 \pm 0.01$ & $0.52\pm 0.02$  \\
$TR_3$ & $0.54 \pm 0.01$ & $0.52 \pm 0.02$    \\
$TR_4$ & $0.53 \pm 0.01$ & $0.52 \pm 0.02$    \\
$TR_5$ & $0.52 \pm 0.01$ & $0.50 \pm 0.01$    \\
$TR_6$ & $0.51 \pm 0.02$ & $0.49 \pm 0.01$   \\
$TR_7$ & $0.49 \pm 0.02$ & $0.48 \pm 0.01$     \\
$TR_8$ & $0.52 \pm 0.01$ & $0.50 \pm 0.02$     \\
$TR_9$ & $0.51 \pm 0.01$ & $0.48 \pm 0.01$    \\
$TR_{10}$ & $0.51 \pm 0.01$ & $0.48 \pm 0.02$       \\
$TR_{11}$ & $0.56 \pm 0.01$ & $0.54 \pm 0.02$   \\
$TR_{12}$ & $0.55 \pm 0.01$ & $0.53 \pm 0.02$   \\
\end{tabular}
\end{table}
\newpage

\begin{table}
\caption{The averaged $\beta$ values computed from the power spectra of the Foreign Exchange rates, Stock Market indices and Treasury rates in the time period from $1997$ to $2001$.}
\label{t.betatimeperiod9701}
\begin{tabular}{cccccccccccc}
$ Data $ & Averaged $\beta$ & $ Data $ & Averaged $\beta$\\
\hline
Foreign Exchange rates \\
HKD & $1.6 \pm 0.2$ & GBP & $1.79 \pm 0.03$    \\
ITL & $1.80 \pm 0.03$ & FRF & $1.81 \pm 0.04$ \\
PHP & $1.8 \pm 0.1$ & PLN & $1.79 \pm 0.04$    \\
AUD & $1.8 \pm 0.1$ & PEN & $1.6 \pm 0.2$    \\
NZD & $1.8 \pm 0.1$ & TRL & $1.7 \pm 0.1$    \\
ILS & $1.8 \pm 0.1$ & THB & $1.83 \pm 0.03$   \\
CAD & $1.80 \pm 0.03$ & PESO & $1.81 \pm 0.04$    \\
SGD & $1.81 \pm 0.02$ & MYR & $1.8 \pm 0.1$    \\
NLG & $1.81 \pm 0.04$ & INR & $1.8 \pm 0.1$   \\
JPY & $1.9 \pm 0.1$ & IDR & $1.83 \pm 0.04$    \\
ESP & $1.80 \pm 0.04$ & TWD & $1.8 \pm 0.1$ \\
KRW & $1.8 \pm 0.1$ & RUB & $2.1 \pm 0.3$    \\
HUF & $1.80 \pm 0.03$ & VEB & $1.8 \pm 0.1$   \\
DEM & $1.81 \pm 0.03$ & BRA & $2.0 \pm 0.2$   \\
CHF & $1.8 \pm 0.1$ \\
\end{tabular}
\end{table}

\begin{table}
\begin{tabular}{cccccccccccc}
Table $7$ (continued)\\
$ Data $ & Averaged $\beta$ & $ Data $ & Averaged $\beta$\\
\hline
Stock Market indices\\
Nasdaq 100 & $1.7 \pm 0.1$ & TW & $1.9 \pm 0.1$    \\
S\&P 500 & $1.8 \pm 0.1$ & ME & $1.8 \pm 0.1$    \\
Nikkei 225 & $1.8 \pm 0.1$ & HS & $1.8 \pm 0.1$    \\
DJIA & $1.80 \pm 0.03$ & BSES & $1.82 \pm 0.03$    \\
CAC 40 & $1.8 \pm 0.1$ & BO & $1.80 \pm 0.02$    \\
AO & $1.8 \pm 0.1$ & MSE & $1.9 \pm 0.1$    \\
FTSE 100 & $1.81 \pm 0.03$ & ASS & $1.9 \pm 0.1$    \\
AEX & $1.8 \pm 0.1$ & BUX & $1.82 \pm 0.04$    \\
DAX & $1.8 \pm 0.1$ & WIG & $1.8 \pm 0.1$    \\
SM & $1.8 \pm 0.1$ & KLSEC & $1.8 \pm 0.1$    \\

T30C & $1.8 \pm 0.1$ & BSET & $1.9 \pm 0.1$    \\
T25 & $1.9 \pm 0.1$ & CO & $2.0 \pm 0.2$    \\
SC & $1.9 \pm 0.1$ & ICB & $2.0 \pm 0.2$    \\
SE 100 & $1.9 \pm 0.1$ & LSEG & $2.0 \pm 0.2$    \\
BCI 30 & $1.9 \pm 0.1$ & JSXC & $1.9 \pm 0.1$    \\
IBEX 35 & $1.8 \pm 0.1$ & AK\&M & $1.9 \pm 0.2$    \\

\end{tabular}
\end{table}

\begin{table}
\begin{tabular}{cccccccccccc}
Table $7$ (continued)\\
$ Data $ & Averaged $\beta$ & $ Data $ & Averaged $\beta$\\
\hline
Treasury rates\\
$TR_1$ & $1.8 \pm 0.1$ & $TR_7$ & $1.9 \pm 0.1$    \\
$TR_2$ & $1.83 \pm 0.04$ & $TR_8$ & $1.9 \pm 0.1$    \\
$TR_3$ & $1.86 \pm 0.05$ & $TR_9$ & $1.8 \pm 0.1$    \\
$TR_4$ & $1.88 \pm 0.06$ & $TR_{10}$ & $1.82 \pm 0.04$    \\
$TR_5$ & $1.9 \pm 0.1$ & $TR_{11}$ & $1.85 \pm 0.04$    \\
$TR_6$ & $1.9 \pm 0.1$ & $TR_{12}$ & $1.9 \pm 0.1$    \\
\end{tabular}
\end{table}

\end{document}